\documentclass[preprints,technicalnote,accept,pdftex,moreauthors]{mdpi}

\newcommand\Teff{\ensuremath{T_\mathrm{eff}}}
\newcommand\logg{\ensuremath{\log g}}
\newcommand\vsini{\ensuremath{v_{e}\sin i}}

\newcommand\met{\ensuremath{[M/H]}}
\newcommand\micro{\ensuremath{\xi_t}}
\firstpage{1} 
\pubvolume{1}
\issuenum{1}
\articlenumber{0}
\pubyear{2024}
\copyrightyear{2024}
\datereceived{ } 
\daterevised{ } 
\dateaccepted{ } 
\datepublished{ } 
\hreflink{https://doi.org/} 

\Title{Generating stellar spectra using Neural Networks}

\TitleCitation{Title}

\Author{Marwan Gebran $^{1,*}$\orcidA{0000-0002-8675-4000}}

\AuthorCitation{Gebran, M.}

\address{%
$^{1}$ \quad Affiliation 1; mgebrane@saintmarys.edu}

\corres{Correspondence: mgebrane@saintmarys.edu}

\firstnote{Current address: Department of Chemistry and Physics, Saint Mary’s College, Notre Dame, IN 46556, USA} 

\abstract{A new generative technique is presented in this paper that uses Deep Learning to reconstruct stellar spectra based on a set of stellar parameters.
Two different Neural Networks were trained allowing the generation of new spectra. First, an autoencoder is trained on a set of BAFGK synthetic data calculated using \texttt{ATLAS9} model atmospheres and \texttt{SYNSPEC} radiative transfer code. These spectra are calculated in the wavelength range of Gaia RVS between 8\,400 and 8\,800 \AA. Second, we trained a Fully Dense Neural Network to relate the stellar parameters to the Latent Space of the autoencoder. Finally, we linked the Fully Dense Neural Network to the decoder part of the autoencoder and we built a model that uses as input any combination of \Teff, \logg, \vsini, \met, and \micro \ and output a normalized spectrum.
The generated spectra are shown to represent all the line profiles and flux values as the ones calculated using the classical radiative transfer code. The accuracy of our technique is tested using a stellar parameter determination procedure and the results show that the generated spectra have the same characteristics as the synthetic ones.}


\begin{document}


\section{Introduction}

Most of the stellar spectroscopy analysis techniques require synthetic data in order to be tested and constrained (Kassounian et al., 2019; Gebran et al.; 2016, 2022, 2023; Gilda, 2023). Astronomers have used radiative transfer codes to simulate the spectra of specific stars, planets, galaxies, and other astronomical objects. Synthetic stellar spectroscopy relies on few combination of model atmospheres and radiative transfer codes.  It is worth mentioning the \texttt{MARCS/Turbospectrcum} models (Gustafsson et al., 2008; Plez, 2012) that are mainly used for giants and dwarfs stars. \texttt{MOOG} code (sneden et al., 2012) is able to perform a variety of LTE line analysis and spectrum synthesis tasks.  \texttt{ATLAS/SYNTHE/WIDTH} (Kurucz, 1993; Sbordone et al., 2004) is used for spectral synthesis and the derivation of chemical abundances from equivalent widths of spectral lines. \texttt{SME} (Piskunov \& Valenti, 2017) is capable of deriving stellar parameters and chemical composition for a broad range of stars, including cool dwarfs and giants. \texttt{SYNSPEC} (Hubeny \& Lanz, 2011, 2017; Hubeny et al., 2021), used in this work, can synthesize spectra of stars with effective temperatures (\Teff) $\geq$4\,000 K. Finally, \texttt{PHOENIX} models (Husser et al., 2013) are well suited for stars having \Teff$\leq$12\,000 K.\\

When users do not have direct access to these codes and wrappers, they use data from available online databases. These databases are usually calculated with large steps in \Teff \ and surface gravity, \logg, leading to large uncertainties on the derived stellar parameters once compared with true observations. Examples of these databases are \texttt{POLLUX} (Palacios et al., 2010) which contains spectra of stars varying in \Teff, \logg, and metallicity (\met). \texttt{TLUSTY} is a Non-LTE Line-blanketed model atmospheres of O-type stars (Lanz \& Hubeny, 2003) with \Teff$\geq$ 27\,500 K. 
 For cooler stars, the \texttt{AMBRE} database contains high-resolution FGKM synthetic spectra (de Laverny et al., 2012).\\

In this work, we suggest a new technique based on Artificial Neural Networks (ANN) that learns from the relation between stellar parameters and the flux of synthetic data and then allows the user to calculate a synthetic spectrum by providing the ANN a combination of \Teff, \logg, \met, projected equatorial rotational velocity (\vsini), microturbulence velocity (\micro), and resolution.
The advantage of such a technique is that it only requires to have Python installed and no need to have any compiler such as \texttt{Fortran} or \texttt{C++} nor a radiative transfer code. Ultimately, one would use this technique to train an ANN using well-known observed spectra. The spectra should have a large range of wavelength, stellar parameters, and individual chemical abundances. This can only be done by combining the data from different surveys to increase the wavelength range of the spectra. 
Being able to reproduce stellar spectra without the use of model atmospheres and radiative transfer code will make spectral synthesis homogeneous across the HR-Diagram. It will also allow the comparison of results among different studies as we won't be faced with the classical issue of using different physics in different types of models (e.g. MARCS vs. ATLAS or Turbospectrcum vs. SYNSPEC).

For the sake of the proof of concept, this paper will detail the technique when applied to a small wavelength range. We have decided to chose the range of Gaia Radial Velocity Spectrometer (RVS, Cropper et al. 2018) based on the fact that it has more than 1 million mean spectra (Gaia Collaboration et al. 2023) and that it contains a small wavelength range ($\lambda$ between 8\,400 and 8\,800 \AA), making the calculation of the training database less time consuming. The RVS will provide spectra in the CaII IR triplet region combined with weak lines of Si I, Ti I, Fe I, among others. Weak lines of N I, S I, and Si I and strong Paschen hydrogen lines can be found in A and F stars. For stars hotter than A0, spectra may contain lines of N I, Al II, Ne I, Fe II, and He I (Recio-Blanco et al., 2016). RVS spectra contain valuable information about the stellar parameters as will be discussed in this paper.

Section \ref{method} will explain the construction of the synthetic database and the Neural Networks (NNs) used in this work. Section \ref{results} will show the newly generated spectra that will be tested against a stellar parameters derivation technique. Finally, the discussion and conclusions will be presented in Sec.~\ref{discussion}.

\section{Materials and Methods}
\label{method}
The technique of generating synthetic spectra requires several steps. The first one is to calculate the database of synthetic data, the second step requires building autoencoders in order to reduce the dimension of the original flux to fewer data points that are saved in a Latent Space, and finally, the fundamental parameters are related to the Latent Space through a fully connected NN. Conceptually, we will developing a model that takes as an input the stellar parameters and generates as an output a synthetic spectrum. 
An extra step was performed in Sec.~\ref{results} for deriving the stellar parameters of the generated dataset. This is done in order to check the reliability and accuracy of the generated data.

\subsection{Training Database}
The details of calculating a training database are fully explained in Gebran et al. (2023). In summary, Line-blanketed model atmospheres were calculated for the purpose of this work using \texttt{ATLAS9} (Kurucz, 1992). These are LTE plane parallel models that assume hydrostatic and radiative equilibrium. We have used the Opacity
Distribution Function (ODF) of Castelli \& Kurucz (2003). Using  Smalley’s (2004) prescriptions, we have included convection in the atmospheres of stars cooler than 8500 K. We have treated convection using the mixing length theory. A mixing length parameter of 0.5 was used for 7\,000 K$\leq$\Teff$\leq$8500 K, and 1.25 for \Teff$\leq$7\,000 K. The synthetic spectra grid was computed using \texttt{SYNSPEC} (Hubeny \& Lanz, 2011) according to the parameters described in Tab.~\ref{parameters}. Metallicity was scaled, with respect to the Grevesse \& Sauval (1998) solar value, from -1.5 dex up to +1.5 dex. The metallicity is calculated as the abundance of elements heavier than Helium. The change in metallicity consists of a change in the abundance of all metals with the same scaling factor. The synthetic spectra were computed from 8\,400 \AA \ up to 8\,800 \AA \ with a wavelength step of 0.10 \AA. This range is chosen to include the RVS range. The Resolution, ($\frac{\lambda}{\Delta \lambda}$), also varies around the nominal RVS one which is around ~11\,500 (Cropper et al., 2018). As explained in Recio-Blanco et al. (2023), the RVS spectra contain lines with information on the chemical abundance of many metals ( Mg, Si, Ca, Ti, Cr, Fe, Ni, and Zr among others) that have different ionization stages and are sensitive to the stellar parameters. We have used the line list of Gebran et al. (2023).

\begin{table}[!t]
    \centering
    \begin{tabular}{||c|c||}
    \hline
       Parameter  & Range  \\
       \hline
        \Teff & 3\,600 -- 15\,000 K \\
        \logg & 2.0 -- 5.0 dex\\
        \vsini & 0 -- 300 Km/s \\
        \met & -1.5 -- 1.5 dex \\
        \micro & 0 -- 4 Km/s \\
        \hline \hline

        Resolution ($\dfrac{\lambda}{\Delta \lambda}$)& 5000 -- 14500 \\ 
        &\\
        \hline
    \end{tabular}
    \caption{Range of parameters used in the calculation of the synthetic spectra. The upper part of the table displays the astrophysical parameters of the stars whereas the bottom one displays the instrumental one. All these spectra were calculated in a wavelength range of 8\,400 -- 8\,800 \AA. }
    \label{parameters}
\end{table}

\begin{figure}[!h]
    \centering
    \includegraphics[scale=0.44]{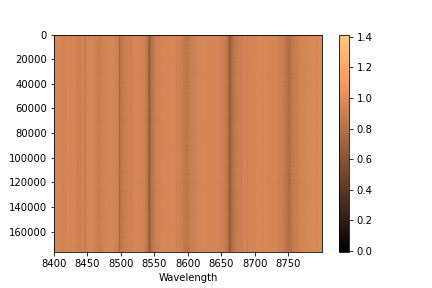}
    \includegraphics[scale=0.18]{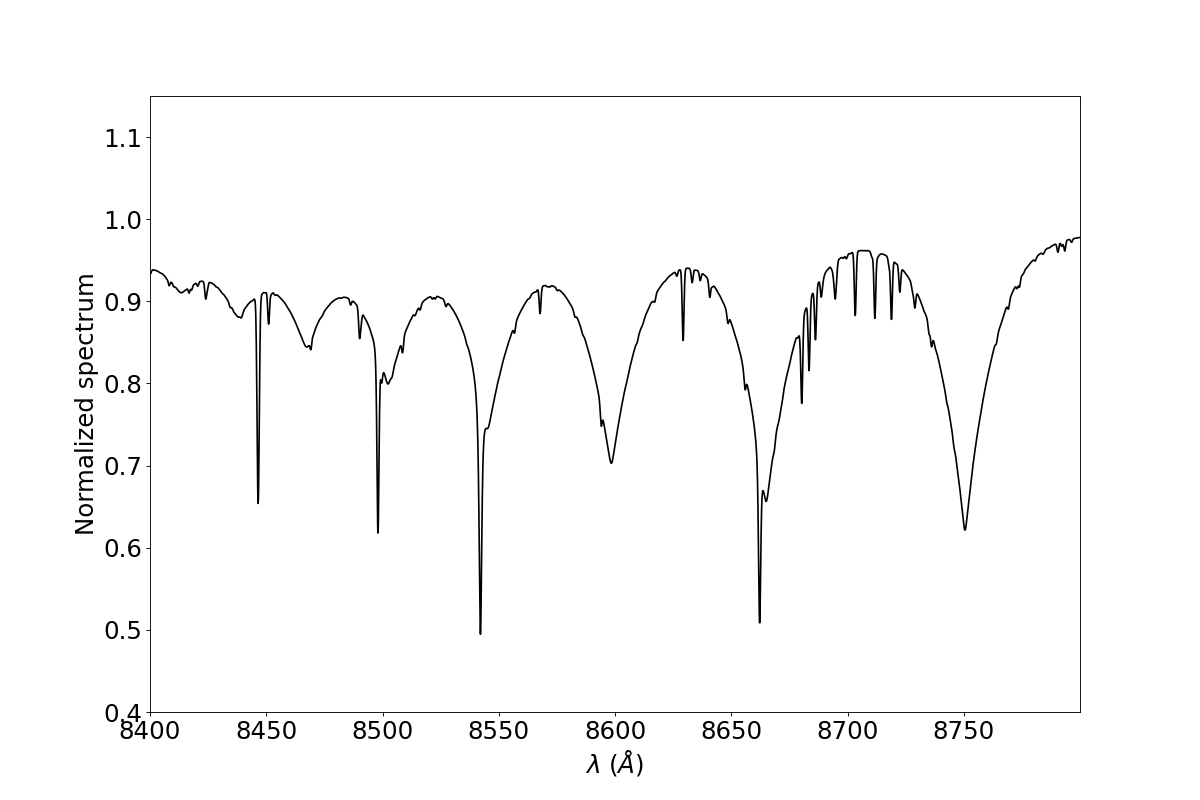}

    \caption{Left: Color map representing the fluxes for a sample
of the training database. Wavelengths are in \AA. Right: Example of a synthetic spectrum without calculated in the RVS wavelength range.}
    \label{colormap}
\end{figure}

 We ended up with a database of 205\,000 synthetic spectra with \Teff, \logg, \vsini, \met, \micro, and resolution chosen randomly from Tab.~\ref{parameters}. The left side of Fig.~\ref{colormap} shows a color map of a sub-sample of the database. The absorption lines of the calcium triplet ($\lambda$= 8498, 8542, 8662 \AA) are easily detected in the figure. The right side of Fig.~\ref{colormap} displays an example of a synthetic spectrum having a \Teff, \logg, \vsini, \met, \micro, and resolution of 9950 K, 3.50 dex, 2 Km/s, 0.5 dex, 1.3 Km/s, and 11500, respectively.

\subsection{Autoencoder}
\label{auto-encoder-part}
Autoencoders, usually used in denoising techniques (Einig et al., 2023; Scourfield et al., 2023), are a type of ANN used in unsupervised machine learning and deep learning. They are primarily used for dimensionality reduction and data compression. Autoencoders usually replace Principal Component Analysis (PCA) because of their non-linear properties. Autoencoders consist of two algorithms, an encoder and a decoder.
They usually work by learning a compact representation of the input data within a lower-dimensional space (Latent Space) and then reconstructing the original data from this representation.\\

All the calculations that are presented in this paper were done using the Machine Learning platform \texttt{TensorFlow}\footnote{\url{https://www.tensorflow.org/}} with the \texttt{Keras}\footnote{\url{https://keras.io/}} interface and both written in \texttt{Python}. We have used a computer that has 52-2.1 GHz cores, Nvidia RTX A6000 48GB graphic card, 256 GB of RAM, and 12 TB of data storage.\\

\begin{figure}[!h]
    \centering
    \includegraphics[scale=0.3]{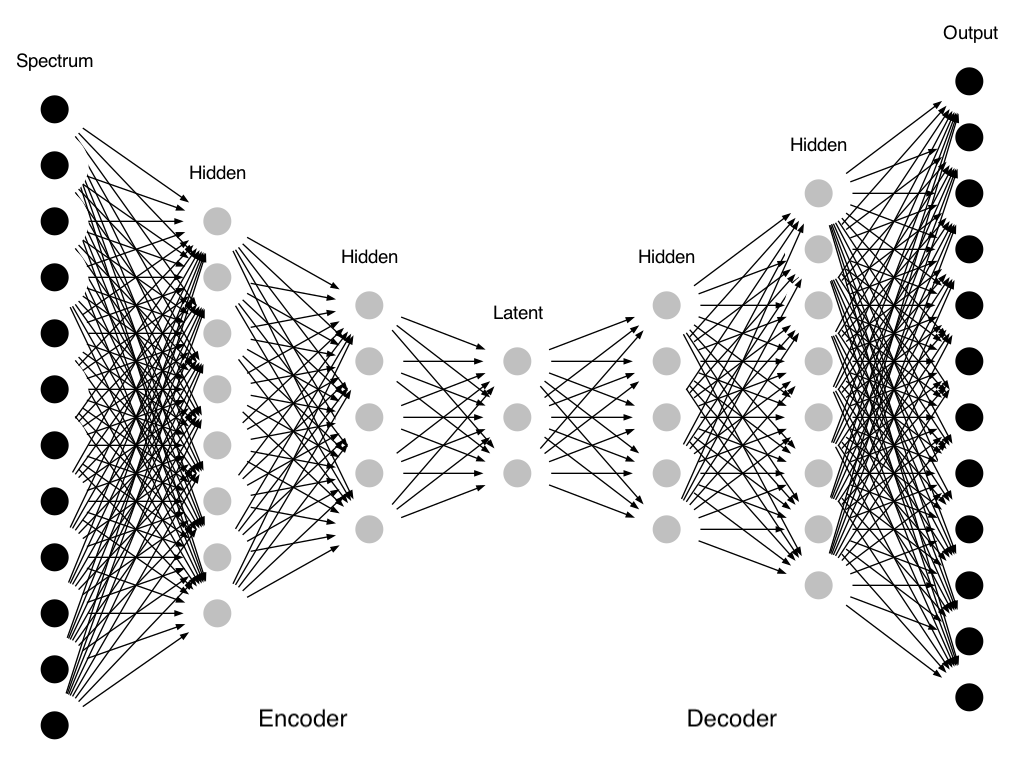}
    \caption{General display of an autoencoder that transforms an input spectrum to a lower dimension using a series of hidden layers and then reconstructs it with its original dimension. The middle layer is the Latent Space that contains the reduced spectrum.   }
    \label{autoencoder}
\end{figure}

The architecture of the autoencoder is shown in Fig.~\ref{autoencoder} in which the initial spectrum is introduced having 4\,000 data points ($\lambda$ between 8\,400 and 8\,800 \AA\ with a step of 0.1 \AA) is then reduced to 10 parameters inside a Latent Space by passing it through a series of hidden layers of different sizes. There is no straightforward rule that relates the size of the Latent Space as a function of the input but we were inspired by the work of Paletou et al. (2015) and Gebran et al. (2016, 2023) in which the authors were able to represent a larger spectral range with only 12 points using PCA. We have used a Latent Space with a dimension of 10 for our spectra. All architectures used in this work as well as the choice of the hyperparameters were constrained using the technique introduced in Gebran et al. (2022). The optimal parameters are the results of minimization procedure of the loss function (see. Gebran et al., 2022, for more details).\\

The steps that transform the spectrum from its original 4\,000 data points to 10 form the encoder part of the autoencoder. A symmetrical counterpart, a decoder, transforms the 10 data points to their original values of 4\,000. The characteristics and parameters used in this autoencoder are presented in Tab.~\ref{auto-param}.\\

\begin{table}[]
    \centering
    \begin{tabular}{||c|c|c||}
    \hline
    Layer & Characteristics & Activation function \\
    \hline
    \multicolumn{3}{||c||}{Encoder}\\
    \hline
     \textbf{Input}    & Spectrum of 4\,000 data points & -- \\
     Hidden    &  1024 neurons & relu \\
      Hidden    &  512 neurons & relu \\
     Hidden    &  256 neurons & relu \\
          Hidden    &  64 neurons & relu \\
     Hidden    &  32 neurons & relu \\
    \textbf{Latent Space}    &  10 neurons & relu \\
    \hline
    \multicolumn{3}{||c||}{Decoder}\\
    \hline
Hidden    &  32 neurons & relu \\
Hidden    &  64 neurons & relu \\
 Hidden    &  256 neurons & relu \\
  Hidden    &  512 neurons & relu \\
  Hidden    &  1024 neurons & relu \\
  \textbf{Output}    & Reconstructed spectrum of 4\,000 data points & -- \\
  \hline
    \end{tabular}
    \caption{Architecture of the autoencoder used in this work.}
    \label{auto-param}
\end{table}

\begin{figure}[!b]
    \centering
    \includegraphics[scale=0.35]{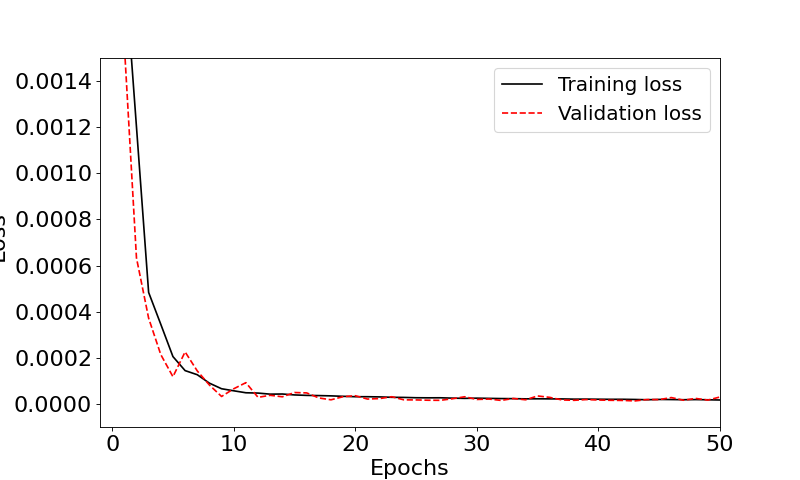}
    \caption{Training and validation loss with respect to the epoch number during the learning process of the autoencoder. Only the first 50 epochs are shown.}
    \label{autoencoder-loss}
\end{figure}

\begin{figure}[!h]
    \centering
    \includegraphics[scale=0.56]{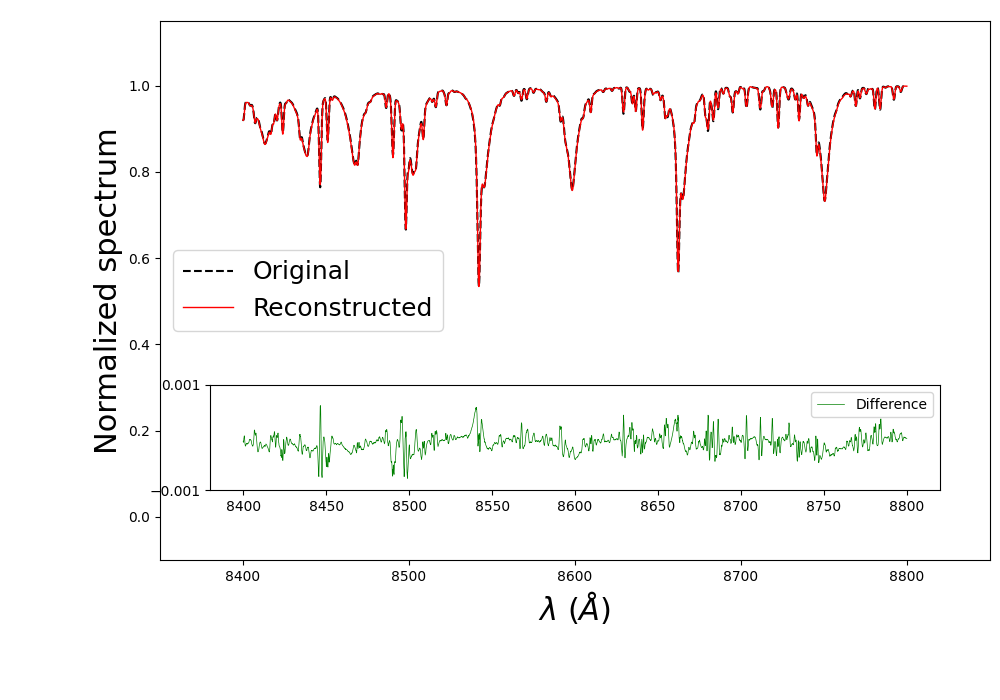}
    \caption{Reconstruction of a synthetic spectrum using the autoencoder of Tab.~\ref{auto-param}. The original spectrum is in black, the reconstructed one is in red, and the difference between them is in green.}
    \label{autoencoder-spec}
\end{figure}

Convergence is set when a monitored metric is stopped improving during training. This stopping criterion is set to be at 15 iterations of non-improvement. Around 75 iterations are required to reach a minimum loss function. Figure~\ref{autoencoder-loss} shows the behavior of the Training and Validation loss as a function of the first 50 epochs. We have used the "Adam" optimizer combined with a Mean Squared Error (MSE) loss function. The precision was verified using different metrics such as the MSE which is found to be $4\times 10^{-6}$, a Mean Absolute Error (MAE) of 0.0013, a variance score of 0.998, and an R$^2$ score of 0.999. A better way to grasp the accuracy is to display an original spectrum along its reconstructed one. Figure~\ref{autoencoder-spec} displays the transformation of a synthetic spectrum that is not included in the training database. The original spectrum (in dashed black lines) passes through the autoencoder in order to be encoded and then decoded. The decoded spectrum is shown in red, and the difference between the original and decoded spectrum is shown in green.  On average, the reconstruction is achieved with an error $<$0.1\%. 

\subsection{Fully Connected Neural Network}
\label{fully-connected-NN}
\begin{figure}[!h]
    \centering
    \includegraphics[scale=0.45]{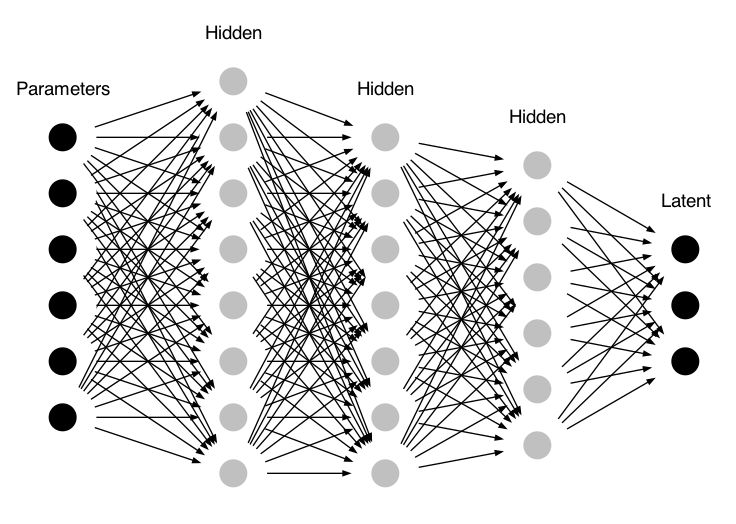}
    \caption{General display of the Fully Connected Neural Network that related the 5 stellar parameters per spectrum and the resolution to the Latent Space.  Six parameters are introduced as input and pass through a series of hidden layers until reaching an output containing the encoded part of the learning database and having the dimension of 10 data points per spectrum.}
    \label{latent}
\end{figure}

The next step of this technique is to link the stellar parameters (\Teff, \logg, \vsini, \met, and \micro), and the resolution of the learning database to the Latent Space through a Fully Connected Neural Network. Using the same database of Sec.~\ref{auto-encoder-part}, We have trained this model by linking the stellar labels of the database to the Latent Space of the same database. This Latent Space is found by passing the spectra database through the encoder part of the autoencoder. This means that the Latent Space is the common the link between the autoencoder and the Fully Connected Neural Network. In the first case, the Latent Space is the output of the encoder part and in the second, the same Latent Space is a transformation of the stellar labels. The architecture of the NN is shown in Fig.~\ref{latent} and the characteristic of each layer is displayed in Tab.~\ref{latent-param}. We have used  "Adamax" as optimizer and an MSE loss function. Input data was divided into batches of 512 spectra per batch. The learning database was divided into 80\% for training and 20\% for validation.

\begin{table}[!h]
    \centering
    \begin{tabular}{||c|c|c||}
    \hline
    Layer & Characteristics & Activation function \\
    \hline

     \textbf{Input}    & Stellar parameters + resolution (6 data points per spectrum) & -- \\
     Hidden    &  5000 neurons & relu \\
     Dropout & 30\% & -- \\
      Hidden    &  2000 neurons & relu \\
      Dropout & 30\% & -- \\
     Hidden    &  512 neurons & relu \\
        Dropout & 30\% & -- \\
  Hidden    &  32 neurons & relu \\
  \textbf{Output}    & Latent Space of 10 data points & -- \\
  \hline
    \end{tabular}
    \caption{Architecture of the Fully Connected Neural Network used to relate the stellar parameters and resolution to the Latent Space.}
    \label{latent-param}
\end{table}

\begin{figure}[!b]
    \centering
    \includegraphics[scale=0.35]{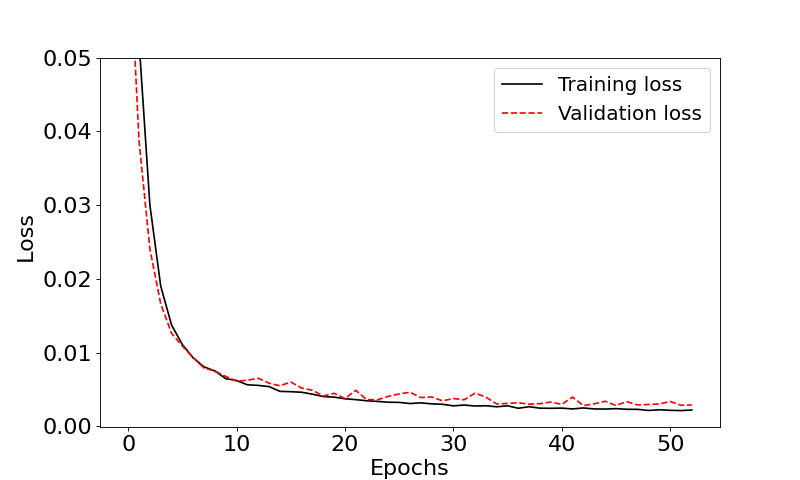}
    \caption{Training and validation loss with respect to the epoch number during the learning process of the Fully Connected Neural Network that related the stellar labels to the Space Space.}
    \label{laten-loss}
\end{figure}

 The training of the model stops when the loss does not decrease significantly after 15 iterations.  The NN requires 50-60 iterations for the loss function to reach a minimum value (i.e. the stopping criterion). This minimum is for both the training and validation databases. Figure~\ref{laten-loss} shows the variation of the Training and Validation loss function as a function of the epochs number. The evaluation of this step is not straightforward. Detailed inspections were performed to make sure that no under- or overfitting is occurring. In the next section (Sec.~\ref{connecting-both}), we will be using this Fully connected network in combination with the decoder of Sec.~\ref{auto-encoder-part} to generate spectra based on input stellar parameters.

\subsection{Connecting both networks}
\label{connecting-both}
The final step is to combine the Fully connected Neural Network of Sec.~\ref{fully-connected-NN} and the decoder part of Sec.~\ref{auto-encoder-part} to generate new data. The main purpose is that the user will input 5 stellar parameters and a resolution as input and get a newly normalized generated spectrum of 4\,000 data points between 8\,400 and 8\,800 \AA.

\begin{figure}[!t]
    \centering
    \includegraphics[scale=0.41]{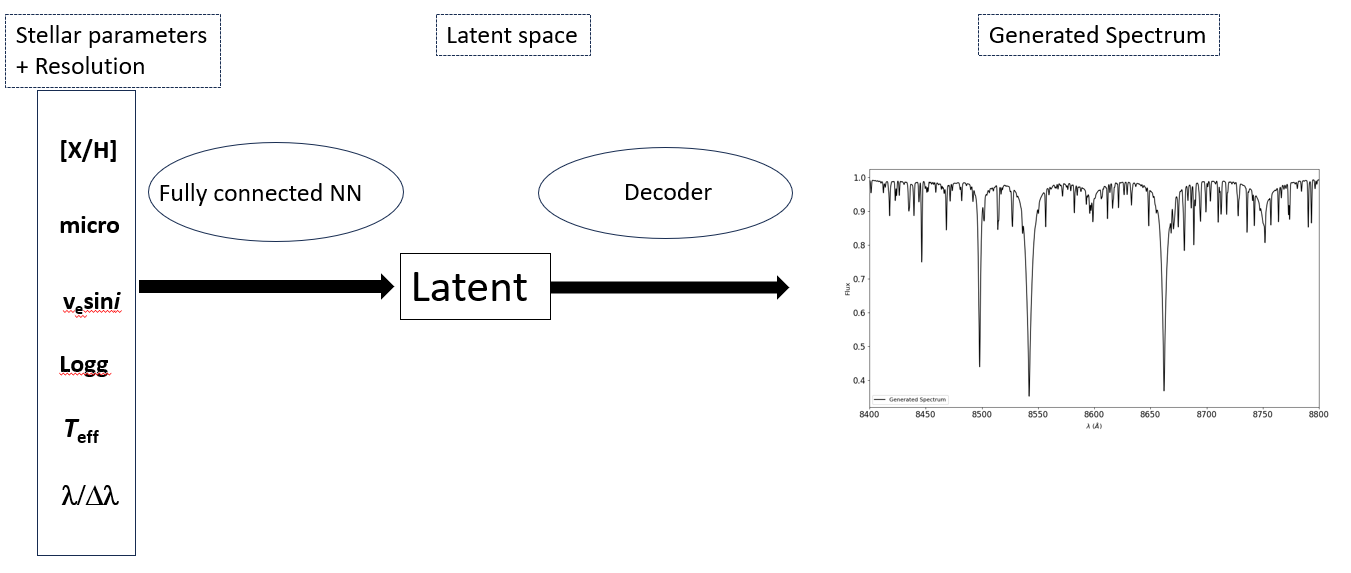}
    \caption{Flowchart of the general steps to generate a synthetic spectrum.}
    \label{flowchart}
\end{figure}

By providing a set of stellar parameters, we can apply the Fully Connected Neural Network of Sec.~\ref{fully-connected-NN} to derive the 10 data points output that will be used a Latent Space. These 10 points are then used as input for the decoder part and decoding them will result in a generated synthetic spectrum of 4\,000  data points. The flowchart of this procedure in shown in Fig.~\ref{flowchart}. Mathematically, the generated synthetic spectrum is calculated as follows:
$$\mathrm{Spectrum}=decoder(NN(\mathrm{Stellar \ Labels}))$$
Where "Stellar Labels" have a dimension of 6, \textit{NN}(Stellar Labels) a dimension of 10, and \textit{decoder}(\textit{NN}(Stellar Labels)) a dimension of 4\,000.
In the next section (Sec.~\ref{results} we will be testing the accuracies of the generated spectra and compare them, through different metrics, to spectra calculated using \texttt{SYNSPEC}.

\section{Results}
\label{results}
Evaluating the quality of the generated spectra is done in two steps. First, The generated spectra were visually inspected along with the ones calculated using \texttt{SYNSPEC}, and second, we performed a detailed stellar parameters analysis of the generated spectra in order to constrain their use in stellar spectroscopy.

\subsection{Generated Spectra}
A set of $\sim$5000 generated spectra was computed using stellar parameters and resolution chosen randomly in the range of Tab.\ref{parameters}. These spectra do not belong to the training database that was used in the autoencoder or the one in the Fully Dense Neural Network.

\begin{figure}[!h]
    \centering
    \includegraphics[scale=0.27]{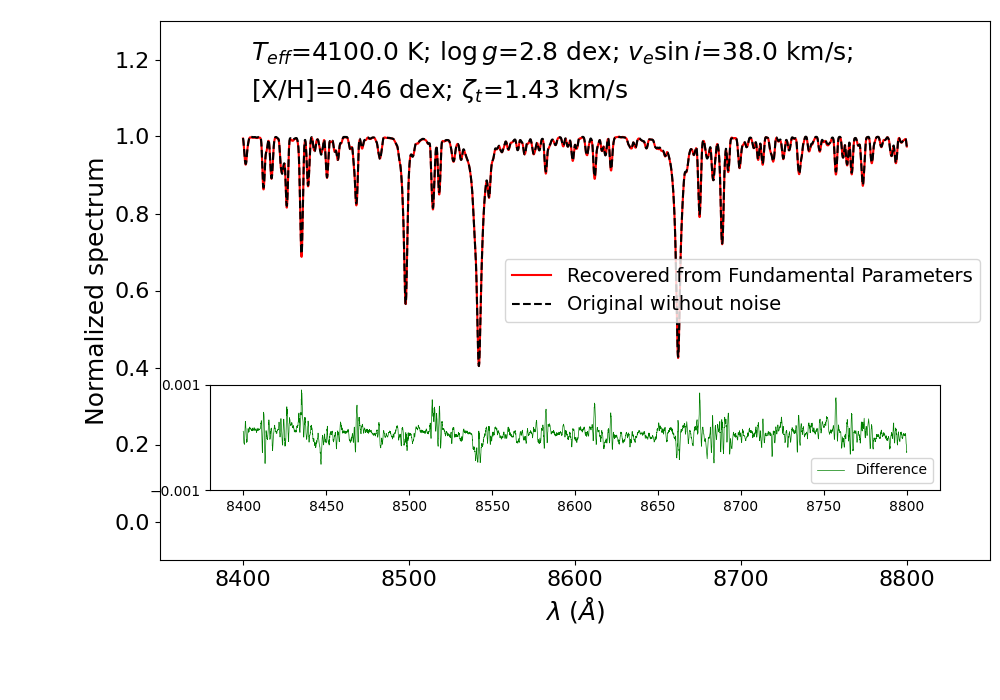}
        \includegraphics[scale=0.27]{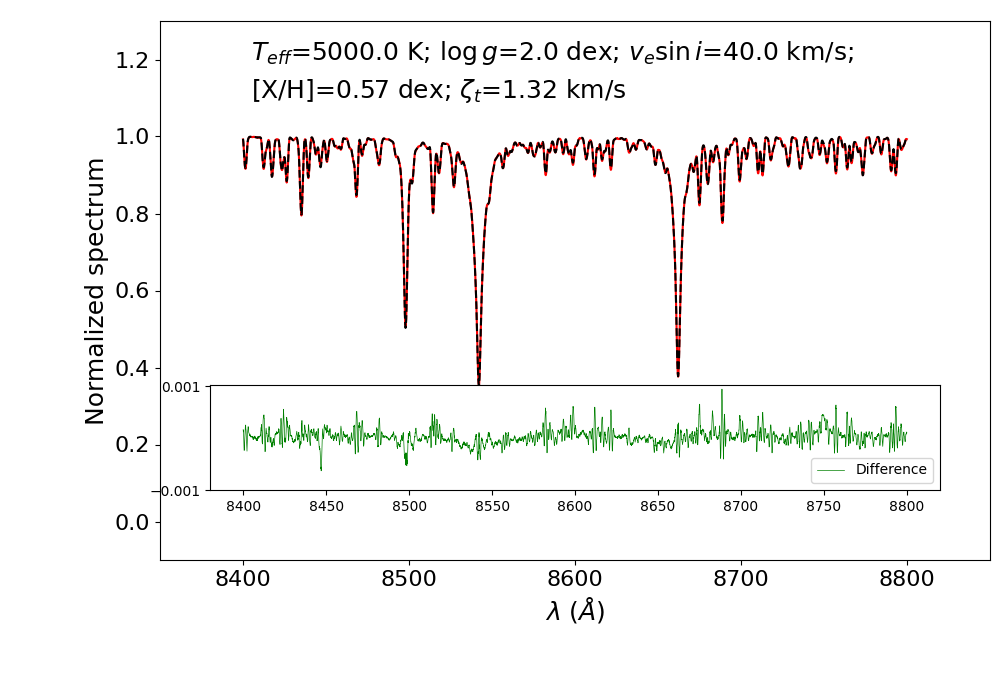}\\
   \includegraphics[scale=0.27]{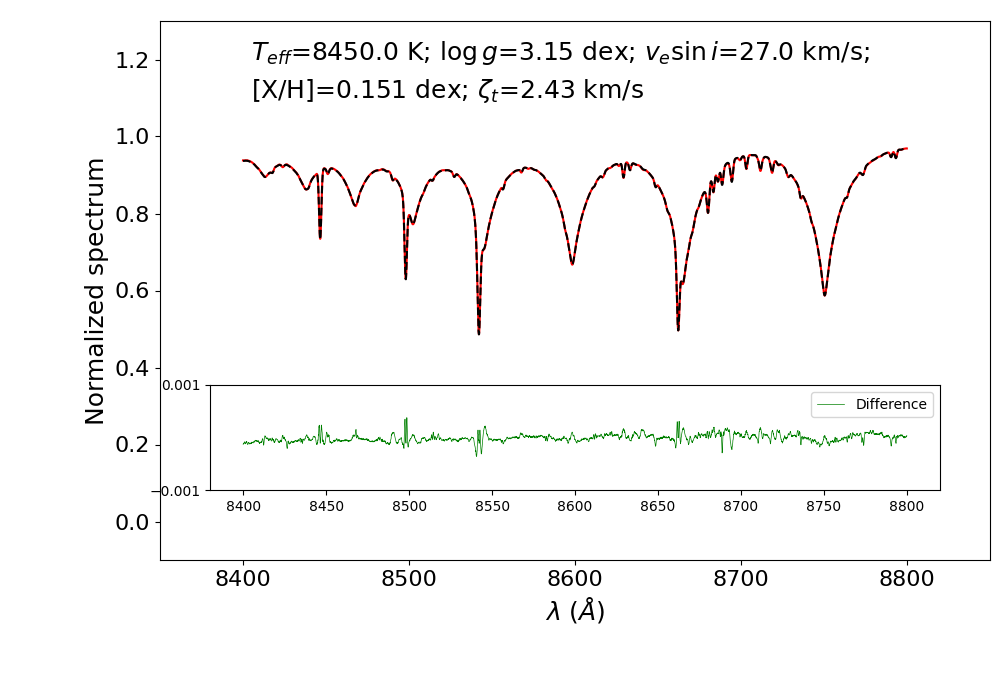}
        \includegraphics[scale=0.27]{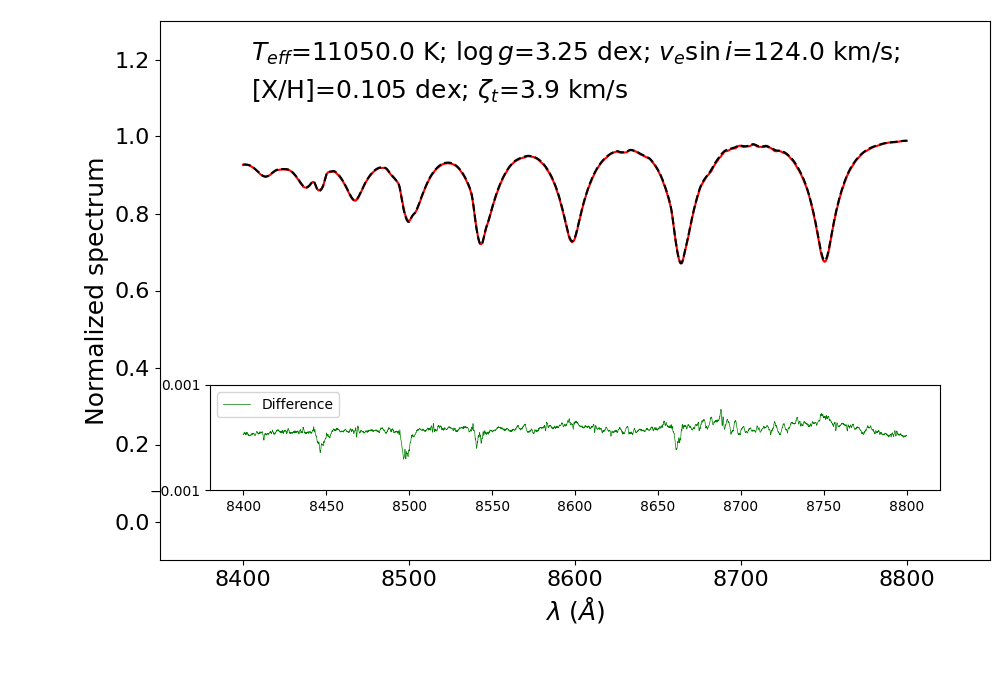}\\
   \includegraphics[scale=0.27]{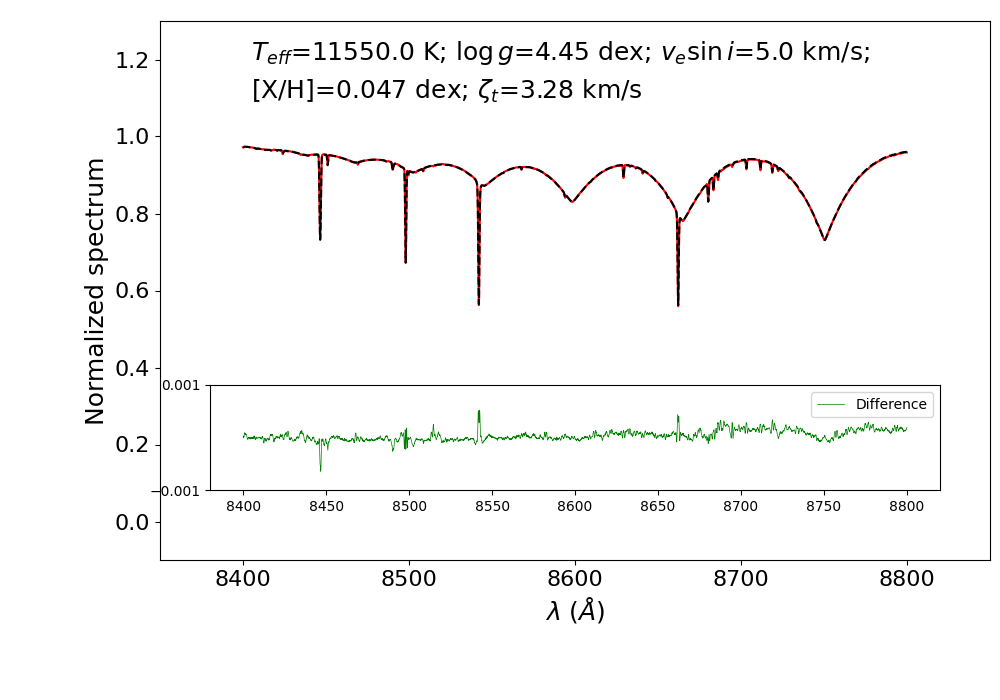}
        \includegraphics[scale=0.27]{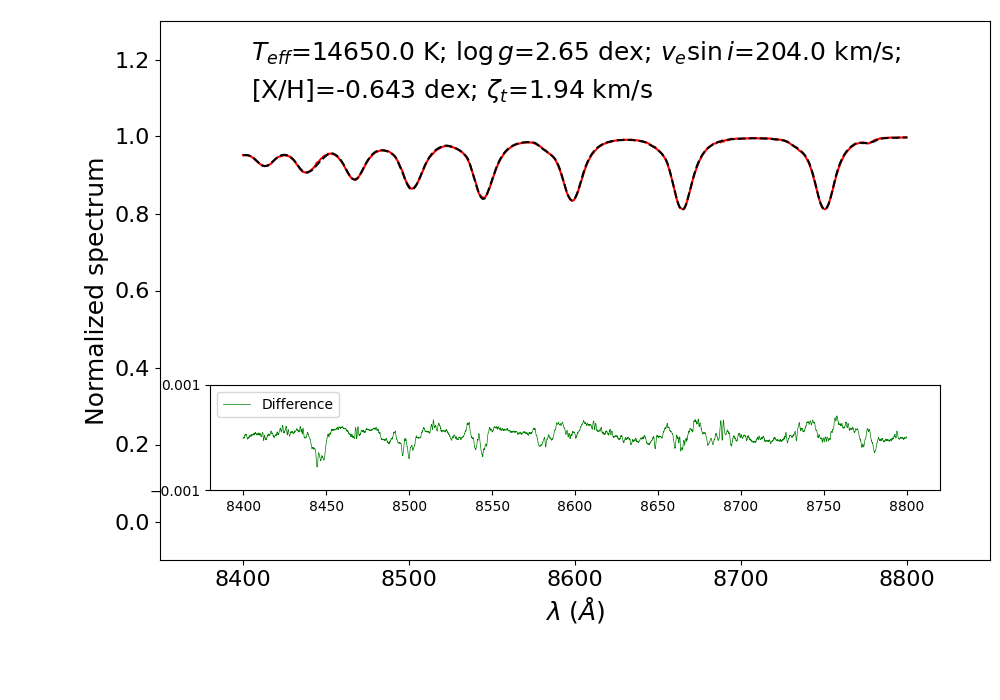}\\
    \caption{Generated spectra (red) compared to the ones calculated using \texttt{SYNEPC} (black) having the same combination of stellar parameters and resolution. The green dashed lines represent the difference between the generated spectra and the synthetic ones. Each spectrum has a different combination of stellar parameters \Teff, \logg, \vsini, \met, and \micro.}
    \label{genvssynspec}
\end{figure}

Qualitatively, the generated spectra are compared to synthetic spectra calculated using \texttt{SYNSPEC} with the same stellar parameters and resolution. Figure.~\ref{genvssynspec} represents a sample of generated plots with each one displaying a generated spectrum (red line) compared to \texttt{SYNSPEC} synthetic spectrum (black dashed line) having both the same stellar parameter and a resolution of 11500. The green dashed lines in the figure represent the difference between the generated and the synthetic spectra. One synthetic spectrum requires a calculation time of $\sim$3 minutes with \texttt{SYNSPEC} whereas a generated spectrum using the combination of NN and decoder requires $\sim$9 ms with the same platform. This step shows that we were able to reconstruct the spectra with detailed line profiles for all combinations of stellar parameters.

\subsection{Stellar parameters}
A quantitative evaluation of the generated spectra is done by inferring the stellar parameters of these spectra. Once the parameters are derived, we compare them to the original parameters used to construct these spectra.  The stellar parameters will be, in that case, derived with a NN trained with \texttt{SYNSPEC} synthetic data. The main goal is to find that the derived stellar parameters have accuracies in the order of the one found when applying the same technique on \texttt{SYNSPEC} synthetic spectra. In that case, the reconstruction generative technique is shown to be able to reproduce the data as if they are calculated using the radiative transfer code.\\

We first start by developing a NN that derives the 6 parameters (\Teff, \logg, \vsini, \met, \micro, and resolution). The network is based on the work of Gebran et al. (2022, 2023) and involves a preprocessing of the original spectra using a PCA transformation. The input data consists of a matrix of synthetic data having a dimension of 205\,000 spectra $\times$ 4\,000 wavelength points. This matrix is reduced to 205\,000 spectra $\times$ 15 PCA coefficients. As explained in Gebran et al. (2023), this step is optional but recommended to increase the speed of the calculations. The choice of the number of coefficients is regulated by the PCA reconstructed error. We found that by retaining only 15 parameters, we were able to reduce the mean reconstructed error of the training database to a value $\leq$0.5\% (see also Paletou et al., 2016 and Gebran et al., 2016, for an explanation about the choice of the number of Principal Components).\\

\begin{table}[!b]
    \centering
    \begin{tabular}{||c|c|c||}
    \hline
    Layer & Characteristics & Activation function \\
    \hline
     \textbf{Input}    & PCA coefficient (15 data points per spectrum) & -- \\
     Hidden    &  5000 neurons & relu \\
      Hidden    &  2000 neurons & relu \\
      Hidden    &  1000 neurons & relu \\
     Hidden    &  512 neurons & relu \\

  Hidden    &  64 neurons & relu \\
  \textbf{Output}    & Stellar Parameters (6 data points per spectrum) & -- \\
  \hline
    \end{tabular}
    \caption{Architecture of the Fully Connected Neural Network used to relate the PCA coefficients to stellar parameters.}
    \label{stellar-param}
\end{table}

The PCA coefficients are used instead of the synthetic spectra in the NN. The NN is then trained to relate these 15 coefficients to the 5 stellar parameters and resolution according to the architecture displayed in Tab.~\ref{stellar-param}. This network is directly adapted from Gebran et al. (2023) and is used for every stellar parameter, independently. The training PCA database is divided in 80\% for training and 20\% for validation. An extra of several tens of thousands of synthetic spectra were calculated for the purpose of being used as test data. The same PCA transformation was used on these test data to derive the 15 coefficients per test spectrum.\\

In this work, we were able to optimize the network for all stellar parameters using a single set of hyperparameters. This means that the network is the same whether we train for \Teff, \logg, or any other parameter. The details about the layers are displayed in Tab.~\ref{stellar-param}. A kernel initializer of "Random Normal", combined with an optimizer of "Adamax" and an "MSE" loss function were used. The input data were divided into batches of 512 spectra (PCA coefficients) per batch. Finally, the loss function reaches a mininum after $\sim$75-200 iterations depending on the stellar parameter.\\

The accuracies on the stellar parameters found in this analysis are shown in Tab.~\ref{Results_param}. These values are calculated as MSE between the original true parameter of the spectra and the ones derived from the NN. Quantitatively, the main result that plays a role in our work is the comparison between the accuracies of the test data and the reconstructed generated data. This comparison shows how accurate the generated spectra using NN/decoder are with respect to the ones calculated using the radiative transfer code \texttt{SYNSPEC}. Of course, none of the sets of test and generated data were used in the training of the NNs. 

\begin{table}[!h]
    \centering
    \begin{tabular}{||c|c|c|c|c||}
         \hline
         Parameter & Training & Validation  & Test  & Generated  \\
         \hline
        \Teff \ (K) & 70  & 110 & 120 & 119 \\
        \logg \ (dex) & 0.02 & 0.03 & 0.03 & 0.05 \\
        \vsini \ (Km/s) & 5.0 & 6.0 & 6.5 & 7.0 \\
        \met \ (dex)&  0.04 & 0.08 & 0.15 & 0.15\\
        \micro \ (Km/s)& 0.15 & 0.17 &  0.17 & 0.20\\
        \hline
    \end{tabular}
    \caption{Derived accuracies of the stellar parameters for the training, validation, test, and generated database.}
    \label{Results_param}
\end{table}

\begin{figure}[!h]
    \centering
    \includegraphics[scale=0.34]{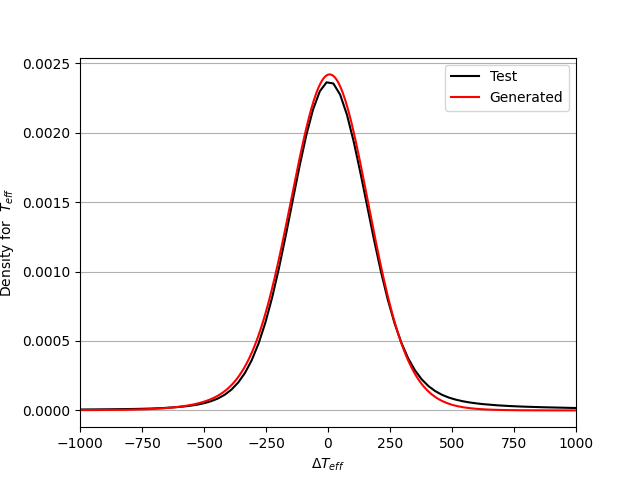}
        \includegraphics[scale=0.34]{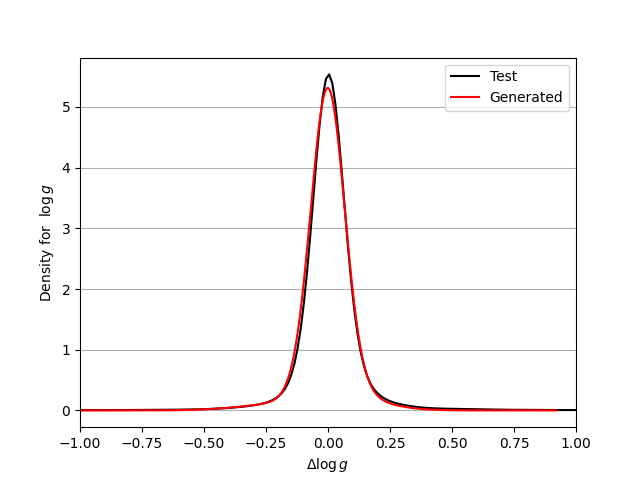}
        \includegraphics[scale=0.34]{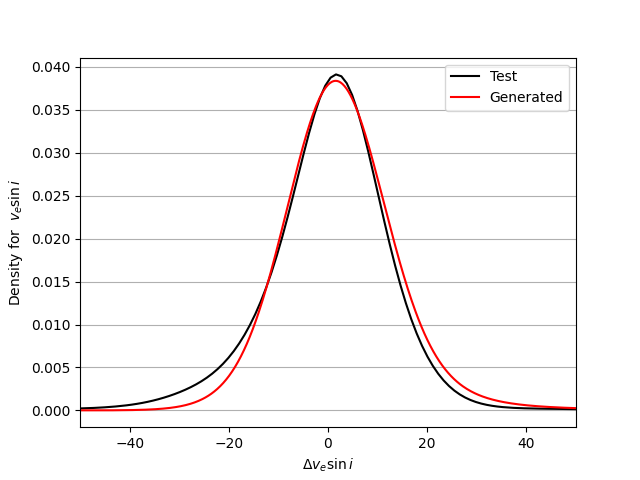}
        \includegraphics[scale=0.34]{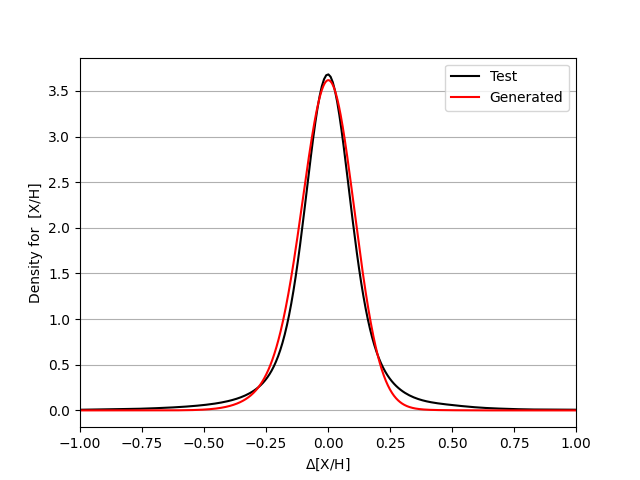}
    \includegraphics[scale=0.34]{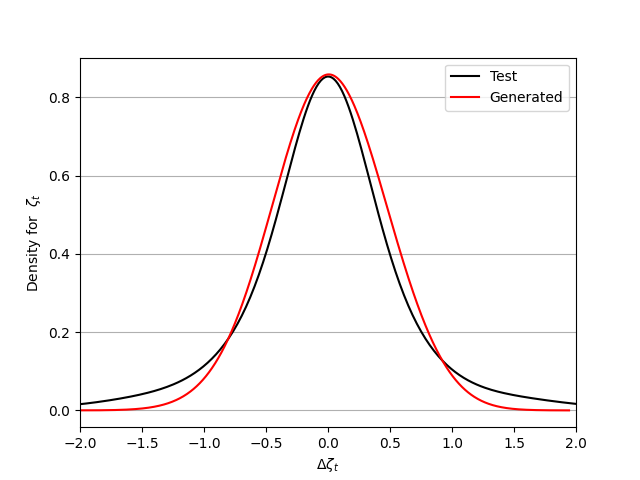}

   \caption{Density distribution of the difference between the predicted values of a stellar label and the true values for the synthetic (black) and generated data (red).}
    \label{density}
\end{figure}

We found that the accuracies on the stellar parameters for the generated spectra (column 5 of Tab.~\ref{Results_param}) are in the same order as the ones of the synthetic spectra calculated with \texttt{SYNCPEC} (Test data, column 4 Tab.~\ref{Results_param}).  These results show that the generative technique is capable of reproducing the flux with all the line profiles and intensities. For a specific stellar parameter, the behavior of the accuracy as a function of the parameter is identical for both synthetic data (\texttt{SYNSPEC}) and generated ones  (NN and decoder). In Fig.~\ref{density}, we represent the density distributions of the difference between the predicted and the true values of \Teff, \logg, \vsini, \met, and \micro, for the synthetic Test (black) and generated data (red). The Generated data behave similarly as synthetic ones, the flux, line profiles, and characteristics are identical to the one calculated using radiative transfer codes.

\section{Discussion and Conclusions}
\label{discussion}
In this work, we have developed a technique, based on deep learning, that takes stellar parameters as input and delivers a generated spectrum as an output. The network is trained using a set of synthetic spectra calculated using the \texttt{SYNSPEC} radiative transfer code and \texttt{ATLAS9} model atmospheres. We were able to generate spectra of BAFGK stars having different surface gravity (\logg), projected equatorial rotational velocity (\vsini), metallicity (\met), microturbulent velocity (\micro), and at a resolution around 11\,500. The choice of the wavelength range (8\,400--8\,800 \AA) and resolution is related to the fact that we were trying to generate RVS spectra-like. We showed that this technique is capable of reproducing the flux and line profiles of stars. Deriving the stellar parameters of the generated spectra shows that the found accuracies are identical to the ones using classical synthetic spectra. \\

In this procedure, we are not constraining the technique of deriving the stellar parameters, rather we are constraining the spectra generation procedure. The astronomical community uses a large number of stellar parameter determination techniques, most of them are based on synthetic data. An example is the GSP-\textit{Spec} (Recio-Blanco et al., 2023) used in the context of Gaia RVS spectra and requires a set of \texttt{MARCS} models and the use of \texttt{TURBOSPECTRUM} radiative transfer code. Our work shows that instead of using synthetic spectra, users can train new models based on the stellar range they need for their project, and generate new stellar spectra to be used in their parameter derivation techniques. The training could be done using spectra from available online databases that are usually calculated with large steps in \Teff and \logg. Such databases are 
\texttt{POLLUX} (Palacios et al., 2010), \texttt{TLUSTY} (Lanz \& Hubeny, 2003), \texttt{PHOENIX} (Husser et al., 2013), and the \texttt{AMBRE} database (de Laverny et al., 2012). Most of these databases contain standard LTE spectra. Training the networks with Non-LTE spectra will result in a more accurate spectra reconstruction and more reliable results when these generated spectra are used to parametrize true observations.\\

Eventually, our models should be trained on observed stars with well-known fundamental parameters (i.e. Benchmark stars). Training the model with observed data and for specific stellar ranges will result in generating a large number of spectra that can be used in the purpose of analysis of large surveys such as RAVE (Steinmetz et al., 2006), the Gaia-ESO Survey (Gilmore et al., 2012), LAMOST (Zhao et al., 2012), APOGEE (Majewski et al., 2017), and GALAH (Martell et al., 2017). For a large wavelength range, we can mention the Melchiors\footnote{\url{https://www.royer.se/melchiors.html}} database (Royer et al., 2023) that combines $\sim$3250 high signal-to-noise spectra of O to M stars between 3\,900 and 9\,000 \AA\ and having a spectral resolution of 85\,000. The generalization of the model presented here is very challenging for observational astronomy. It will require the combination of several surveys with an inspection of every star in terms of the quality of the spectrum and the amount of available stellar labels per star.\\

The next step of this project is to apply the reconstructing technique to a large database containing a wide wavelength range (3\,000--7\,000 \AA), a large range of spectral resolutions ($\dfrac{\lambda}{\Delta \lambda}$ between 1\,000 and 150\,000), and for BAFGKM stars at different evolutionary stages. Having a large spectral resolution will allow the user to generate spectra for a large range of on-ground and space observations. \texttt{SYNSPEC} allows the modifications of individual abundances up to Z=99. We will also be modifying individual chemical abundances instead of the overall metallicity in order to be able to generate spectra with any combination of abundances. Adding to that the range in \vsini\ and \micro, we estimate the size of one database to be in the order of several terabytes. This will require several months of training on our current machines. Once the network is optimized in terms of hyperparameters and architecture, it will be available freely to the scientific community along with a detailed tutorial. Finally, we are currently investigating the relation between the number of spectra in the training database and the wavelength range size. This will allow us to construct databases with optimized number of spectra leading to a significant gain in calculation time.



\vspace{6pt}

\funding{This research received no external funding.}

\acknowledgments{The author acknowledges Saint Mary's College for providing the necessary computational power for the success of this project. The author thanks Drs. Ian Bentley, Fr\'ed\'eric Paletou, and Hikmat Farhat for their helpful comments and suggestions. }


\conflictsofinterest{The authors declare no conflict of interest.} 





\begin{adjustwidth}{-\extralength}{0cm}

\reftitle{References}

\PublishersNote{}
\end{adjustwidth}
\end{document}